\pdfoutput=1
\documentclass[twocolumn,conference]{IEEEtran}
\usepackage[T1]{fontenc}
\usepackage{textcomp}
\usepackage{amsthm}
\usepackage{amsmath}
\usepackage{graphicx}
\usepackage[unicode=true,
 bookmarks=true,bookmarksnumbered=true,bookmarksopen=true,bookmarksopenlevel=1,
 breaklinks=false,pdfborder={0 0 0},backref=false,colorlinks=false]
 {hyperref}
\hypersetup{pdftitle={Your Title},
 pdfauthor={Your Name},
 pdfpagelayout=OneColumn, pdfnewwindow=true, pdfstartview=XYZ, plainpages=false}

\makeatletter

\newcommand{\noun}[1]{\textsc{#1}}

\theoremstyle{plain}
\newtheorem{thm}{\protect\theoremname}
\theoremstyle{plain}
\newtheorem{lem}[thm]{\protect\lemmaname}

\usepackage[caption=false,font=footnotesize]{subfig}

\@ifundefined{showcaptionsetup}{}{%
 \PassOptionsToPackage{caption=false}{subfig}}
\usepackage{subfig}
\makeatother

\providecommand{\lemmaname}{Lemma}
\providecommand{\theoremname}{Theorem}

\begin{document}

\title{Low Complexity Belief Propagation Polar Code Decoder}

\author{\IEEEauthorblockN{Syed Mohsin Abbas, YouZhe Fan, Ji Chen and Chi-Ying Tsui}\IEEEauthorblockA{VLSI Research Laboratory, Department of Electronic and Computer Engineering\\
Hong Kong University of Science and Technology (HKUST), Hong Kong\\
Email: \{smabbas, jasonfan, eejichen, eetsui\}@ust.hk}}
\maketitle
\begin{abstract}
Since their invention, polar codes have received a lot of attention
because of their capacity-achieving performance and low encoding and
decoding complexity. Successive cancellation decoding (SCD) and belief
propagation decoding (BPD) are two approaches for decoding polar codes.
SCD is able to achieve good error-correcting performance and is less
computationally expensive as compared to BPD. However SCD suffers
from long latency due to the serial nature of the successive cancellation
algorithm. BPD is parallel in nature and hence is more attractive
for low latency applications. However, since it is iterative, the
required latency and energy dissipation increases linearly with the
number of iterations. In this work, we borrow the idea of SCD and
propose a novel scheme based on sub-factor-graph freezing to reduce
the average number of computations as well as the average number of
iterations required by BPD, which directly translates into lower latency
and energy dissipation. Simulation results show that the proposed
scheme has no performance degradation and achieves significant reduction
in computation complexity over the existing methods. \end{abstract}


\section{Introduction}

Shanon proved existence of maximum data transmission rate, called
channel capacity \cite{shanon}. Since then, different capacity-approaching
codes have been designed, like Turbo codes \cite{turbo} and LDPC
codes \cite{ldpc}. The first provable capacity-achieving codes, polar
codes, were recently invented by Ar\i kan \cite{arikan}. Polar codes
are considered to be a major breakthrough in coding theory, since
they are the first family of codes known to achieve channel capacity
with explicit construction. Besides achieving the capacity for binary-input
symmetric memoryless channels \cite{arikan}, polar codes were also
proved in \cite{arikan2} to be able to achieve the capacity for any
discrete and continuous memoryless channel. Moreover, an explicit
construction method for polar codes was provided and it was shown
that they can be efficiently encoded and decoded with complexity $\mathcal{O}(n\,log\,n)$,
where $n$ is the code length. Since then, polar codes have become
one of the most popular topics in information theory and have attracted
a lot of attention. 

Several decoding methods are available for decoding polar codes \cite{yazdi}-\cite{20},
SCD and its variants and BPD are two popular methods. SC decoders
suffer from long latency due to the serial nature of the SC algorithm.
However, the SC algorithm requires less computation as compared to
BPD. Based on this property, several high-throughput low-cost SC decoders
were reported in \cite{scdarch}\textendash \cite{22}. Another advantage
of the SC algorithm is its ability to achieve good error-correcting
performance for long code lengths. For short code length, based on
the SCD, the list-decoding or stack decoding method also achieve good
error-correcting performance \cite{list}\textendash \cite{jason2}. 

On the other hand, polar BP decoders \cite{14}-\cite{20} have the
intrinsic advantage of parallel processing. Therefore, compared with
their SC counterparts, polar BP decoders are more attractive for low-latency
applications. For iterative decoders (such as polar BP decoders),
the required latency and energy dissipation increase linearly with
the number of iterations. However, the need for a large number of
iterations makes BP decoders suffer from high computation complexity,
and hence polar BP decoders are still not as attractive as their SC
counterparts. To this end, another decoding method, called soft cancellation
(SCAN) decoding, is proposed in \cite{scan}. By restricting the soft
information propagation schedule in the decoding process, the computational
complexity of SCAN is much lower than that of BPD. However, different
from BPD, the SCAN operation is serial in nature, leading to a much
longer decoding latency. Hence, aiming at the low-latency polar codes
decoder, we concentrate on the BPD in this work.

To address the issues of the large number of iterations and high computation
complexity inherent in BP decoders, Yuan et al. \cite{18} proposed
a G-matrix-based early stopping scheme, which is based on the fact
that iterative decoders normally converge earlier than reaching a
fixed maximum number of iterations. The G-matrix-based stopping criterion
can then be used to stop the computation if convergence has been reached.
To further reduce the computation complexity, in this paper, we propose
a method based on the convergence of the sub-factor-graphs, which
is reached at a much earlier stage. Borrowing the idea from SCD, some
of the sub-factor-graphs are checked during each iteration and if
they have converged, they are frozen and do not need to be computed
in the subsequent iterations. Also the freezing of these sub-factor-graphs
will help to improve the convergence of the decoding process over
rest of the factor graph. As a result, the computation complexity
and also the average number of iterations are reduced. Experimental
results show that our proposed method results in about 40 \textasciitilde{}
46 \% lower computation complexity, as well as lower latency, when
compared to the previously proposed early stopping scheme \cite{18}.

\subsection*{\emph{Notations}}

In this paper, the following notation conventions are used. Matrices
are denoted in boldface capital letters, and vectors in boldface lowercase
letters. The subscript \noun{$_{M}$} of a matrix represents an $MXM$
square matrix and $\mathbf{v}_{M}$ denotes an $MX1$ vector. $x[i]$
stands for the $i^{th}$ element of vector $\mathbf{x}$, $\mathbf{x}^{t}$
stands for vector $\mathbf{x}$ at the $t^{th}$ iteration and $x_{(a:b)}$
represents the sub-vector of $\mathbf{x}$ with the starting and ending
index of $a$ and $b$. The transpose of a vector $\mathbf{x}$ is
denoted by $\mathbf{x}^{T}$. 

\begin{figure}[t]
\includegraphics[width=1\columnwidth]{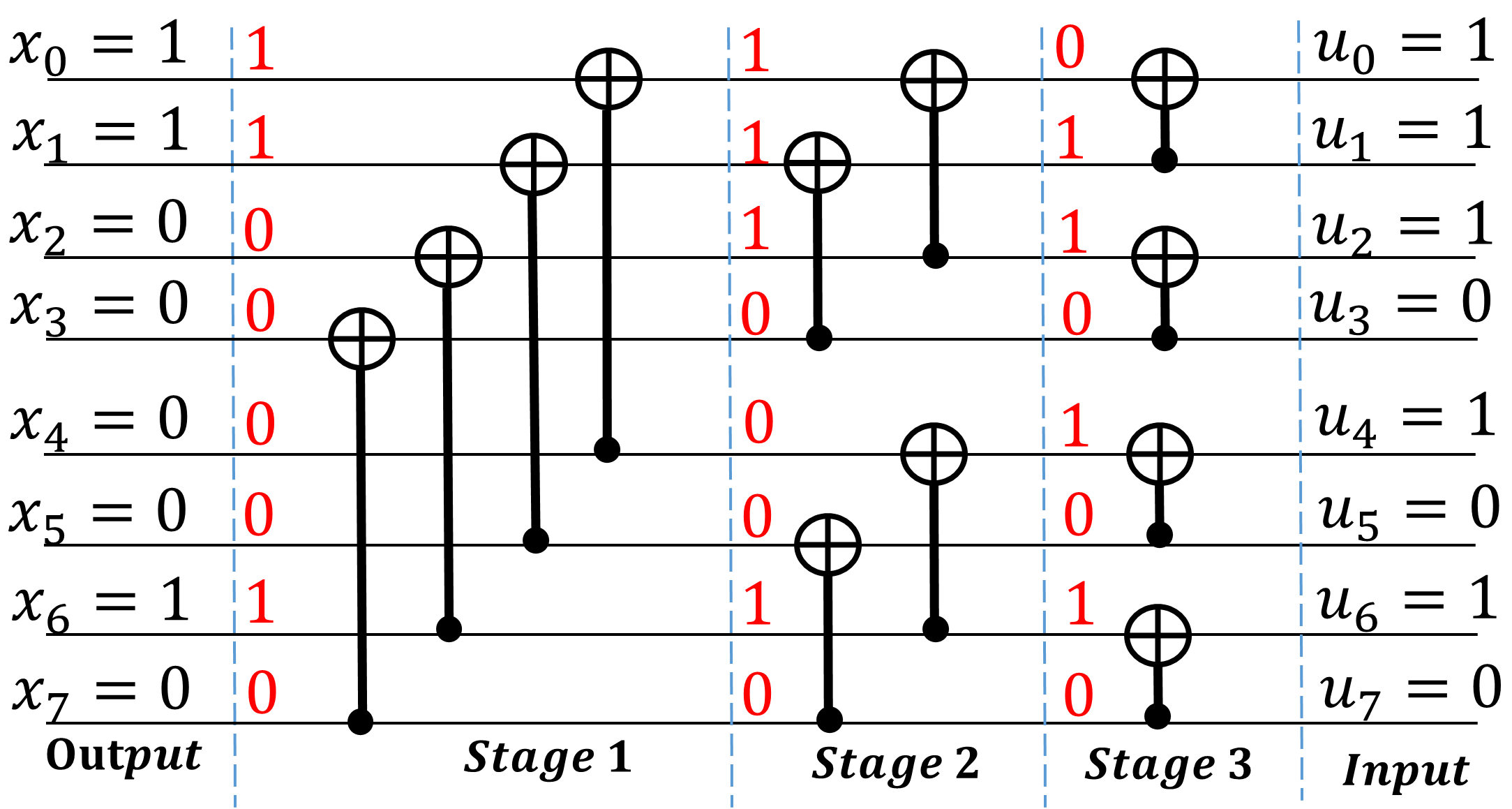}

\protect\caption{Encoding signal flow graph of (8.4) polar code }

\end{figure}

\begin{figure}[t]
\subfloat[]{\protect\includegraphics[width=1\columnwidth]{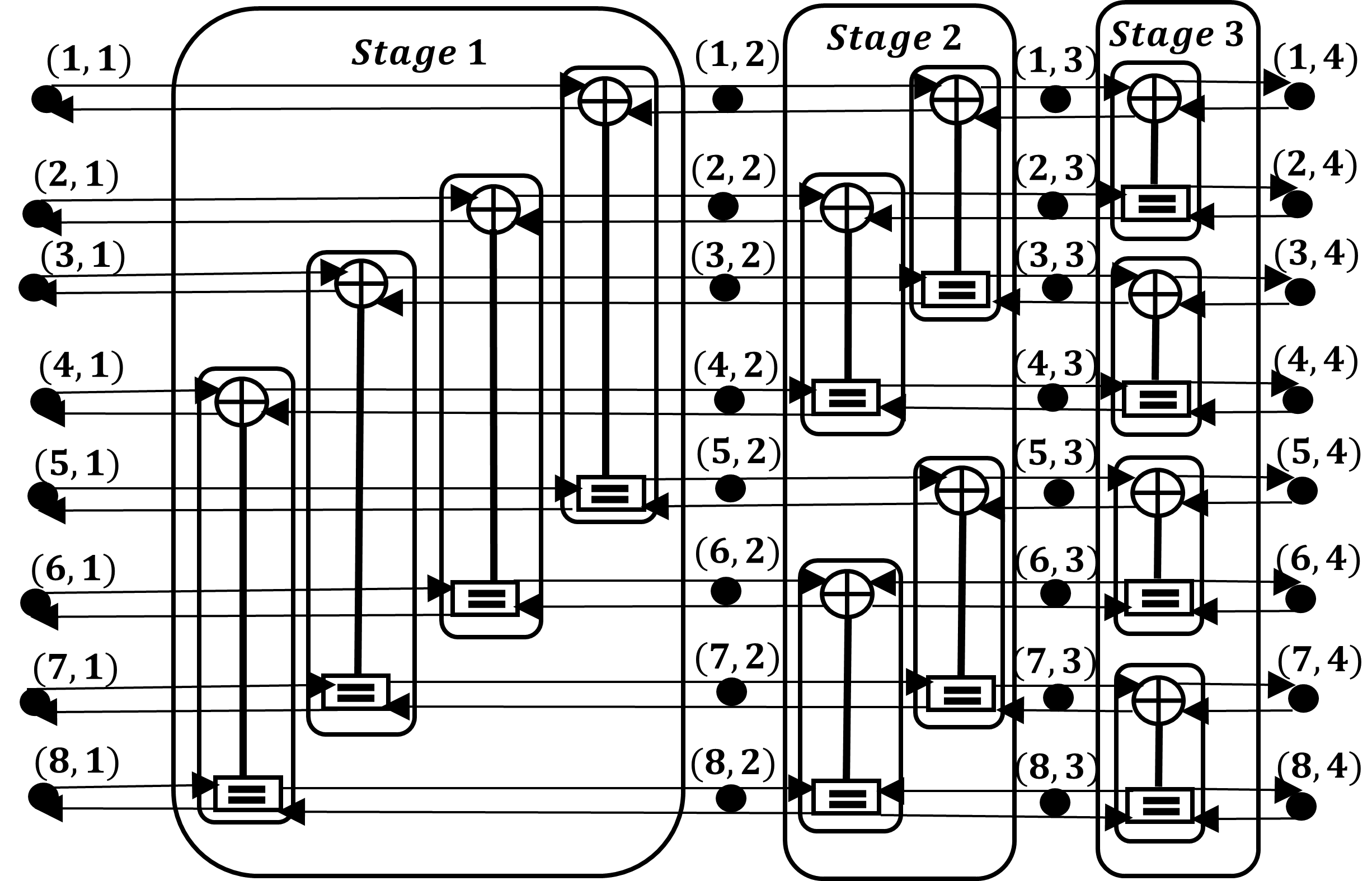}

}

\subfloat[]{\protect\includegraphics[width=0.85\columnwidth]{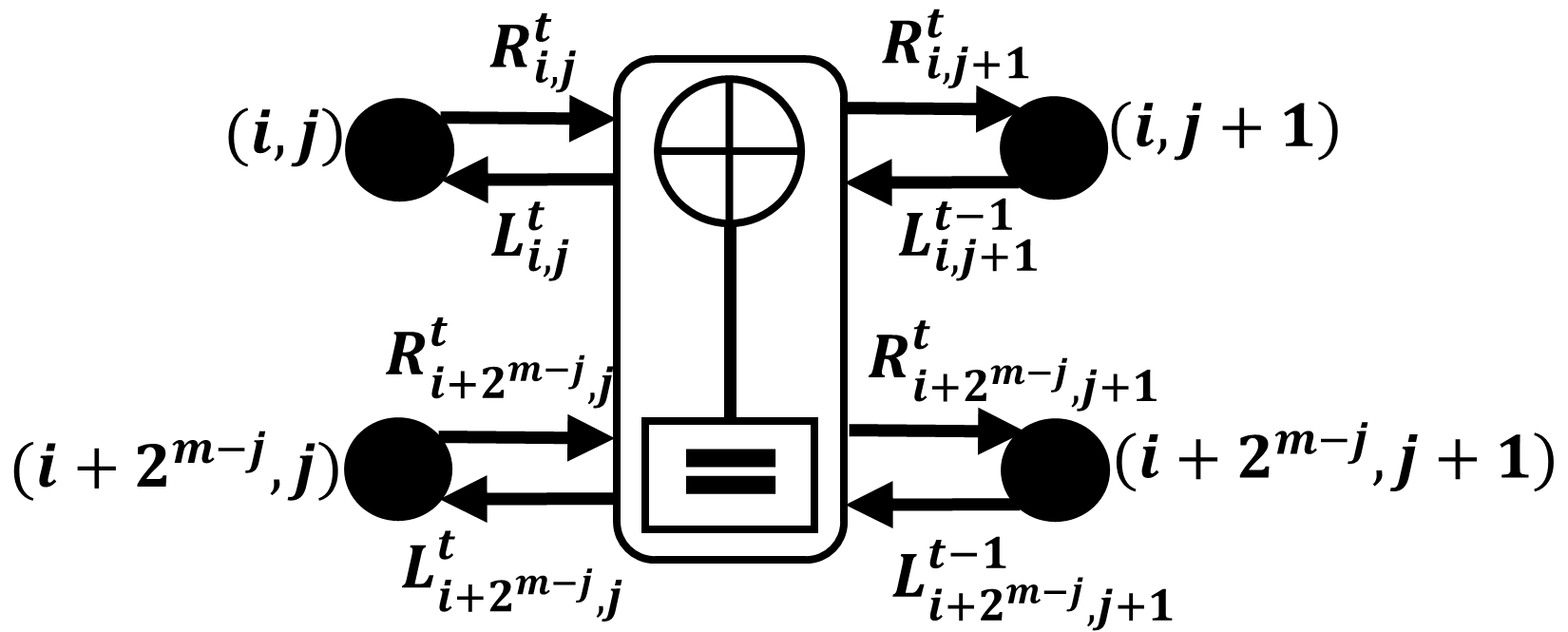}

}

\protect\caption{(a) Factor graph of (8, 4) polar code. (b) Processing Element for
BPD}

\end{figure}

\section{POLAR CODES OVERVIEW}

Polar codes are based on the phenomenon of \textquotedblleft channel
polarization\textquotedblright . More precisely, by recursively combining
and splitting individual channels, some of these channels become essentially
error-free, while others become completely noisy. Furthermore, the
fraction of the noiseless channels tends towards the capacity of the
underlying binary symmetric channels \cite{arikan}. Therefore, an
$(n,k)$ polar code can be generated in two steps. First, an $n$-bit
message $\mathbf{u}$ is constructed by assigning the $k$ reliable
and $(n-k)$ unreliable positions as information bits and \textquotedblleft 0\textquotedblright{}
bits, respectively. The $(n-k)$ unreliable positions, which are forced
to 0, are called the frozen bits (also known as the frozen set $\mathcal{A^{C}}$).
Then, the $n$ -bit $\mathbf{u}$ is multiplied with the generator
matrix $\mathbf{G=F^{\otimes m}}$ to generate an $n$ -bit transmitted
codeword $\mathbf{x}$, where $\mathbf{F}^{\otimes m}$ is the $m^{th}$
Kronecker power of $\mathbf{F}=\left[\begin{array}{cc}
1 & 0\\
1 & 1
\end{array}\right]$ and $m=log{}_{2}n$. Fig. 1 shows the encoding signal flow graph
for $n=8$ polar codes, where the \textquotedblleft $\oplus$\textquotedblright{}
sign represents the XOR operation.

\subsection{Belief Propagation Algorithm for Polar Code Decoding}

As presented in \cite{14}, similar to LDPC codes, polar codes can
be decoded by applying the belief propagation (BP) algorithm over
their factor graphs. For an $(n,k)$ polar code $(n=2^{m})$, the
factor graph is an $m$-stage network consisting of $n.(m+1)$ nodes,
where each node is associated with a right-to-left and a left-to-right
likelihood message denoted by $(L_{i,j}^{t})$ and $(R_{i,j}^{t})$,
respectively. $L_{i,j}^{t}$ denotes the right to left likelihood
message of the $i^{th}$ node at the $j^{th}$ stage and the $t^{th}$
iteration. Fig.2 (a) shows an example of a 3-stage factor graph for
$n=8$ polar codes. Here each stage consists of $n/2=4$ processing
elements (PEs). During the BP decoding procedure, these messages are
propagated and updated among adjacent nodes using the min-sum updating
rule, as shown by the following equations \cite{18}: 

$L_{i+2^{m-j},j}^{t}=L_{i+2^{m-j},j+1}^{t-1}\,+\,\alpha\cdotp\textrm{sign}(L_{i,j+1}^{t-1})\textrm{sign}(R_{i,j}^{t})\textrm{\ensuremath{\cdotp}}\min(|L_{i,j+1}^{t-1}|,|R_{i,j}^{t}|)$

$L_{i,j}^{t}=\alpha\cdotp\textrm{sign}(L_{i,j+1}^{t-1})\textrm{sign}(L_{i+2^{m-j},j+1}^{t-1}+R_{i+2^{m-j},j}^{t})\cdotp\min(|L_{i,j+1}^{t-1}|,|L_{i+2^{m-j},j+1}^{t-1}+R_{i+2^{m-j},j}^{t}|)$

$R_{i,j+1}^{t}=\alpha\cdotp\textrm{sign}(R_{i,j}^{t})\textrm{sign}(L_{i+2^{m-j},j+1}^{t-1}+R_{i+2^{m-j},j}^{t})\cdotp\min(|R_{i,j}^{t}|,|L_{i+2^{m-j},j+1}^{t-1}+R_{i+2^{m-j},j}^{t}|)$

$R_{i+2^{m-j},j+1}^{t}=R_{i+2^{m-j},j}^{t}\,+\,\alpha\cdotp\textrm{sign}(L_{i,j+1}^{t-1})\textrm{sign}(R_{i,j}^{t})\cdotp\min(|L_{i,j+1}^{t-1}|,|R_{i,j}^{t}|)\left.,\left.\left.\left.\left.\right.\right.\right.\right.\right.(1)$ 

$\alpha$ is a scaling parameter introduced in \cite{20} for the
improvement of the decoding performance of a BP decoder. According
to the decoding procedure of BP algorithm, PEs are activated stage-by-stage
from left to right in each iteration. After the number of iterations
reaches the specific maximum number (max\_iter), node $(i,m+1)$ will
output the decoded information bit $u_{i}$ based on the hard decision
of the messages $R_{i,m+1}^{\textrm{max\_iter}}$.

\begin{figure}[t]
\includegraphics[width=1\columnwidth]{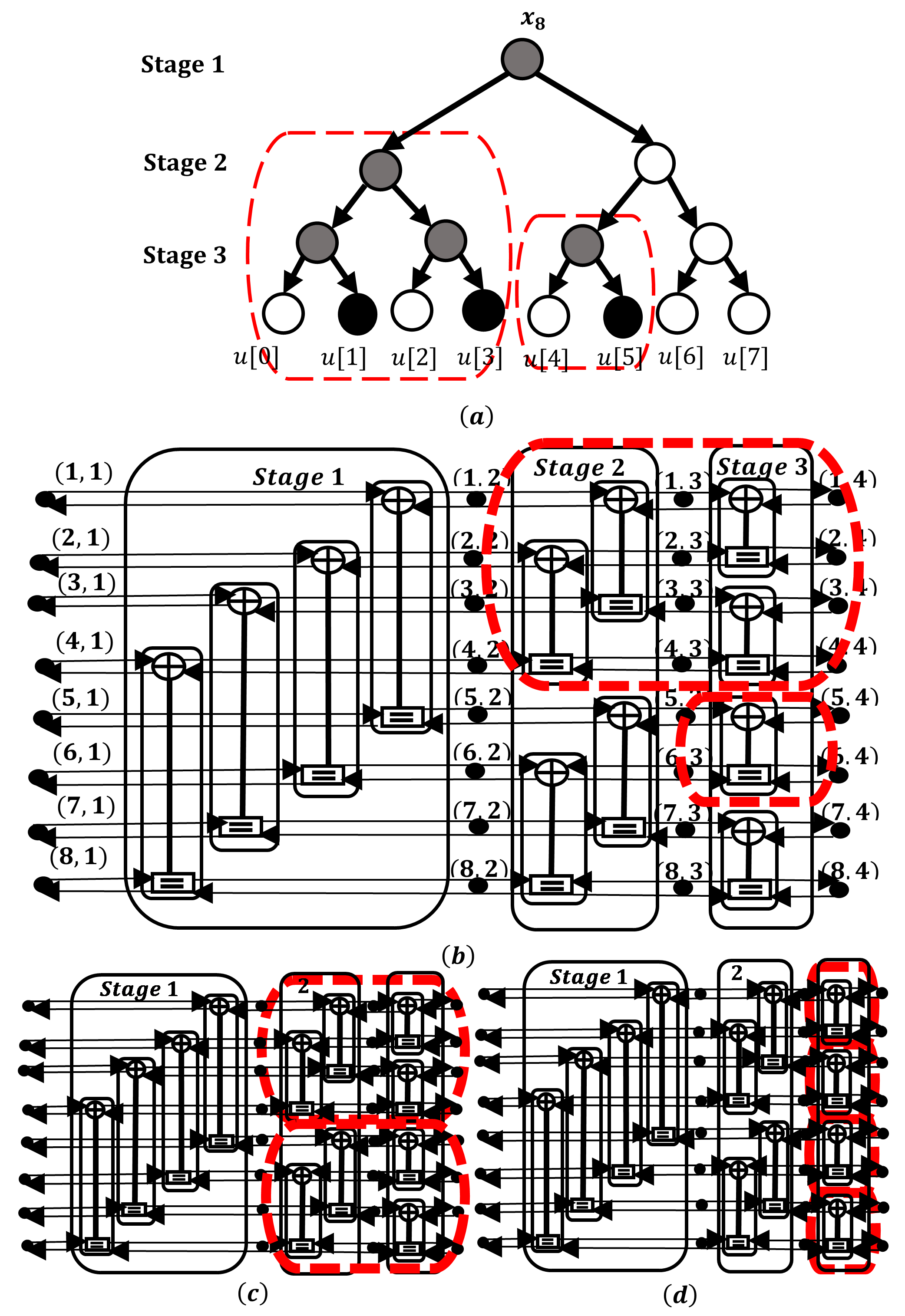}

\protect\caption{Correspondence between SCD scheduling tree and BPD factor graph (a)
SCD Scheduling Tree (b) BPD Factor Graph (c) 2 CSFG's at stage 1 (d)
4 CSFG's at stage 2}

\end{figure}

\begin{figure}[t]
\includegraphics[width=1\columnwidth]{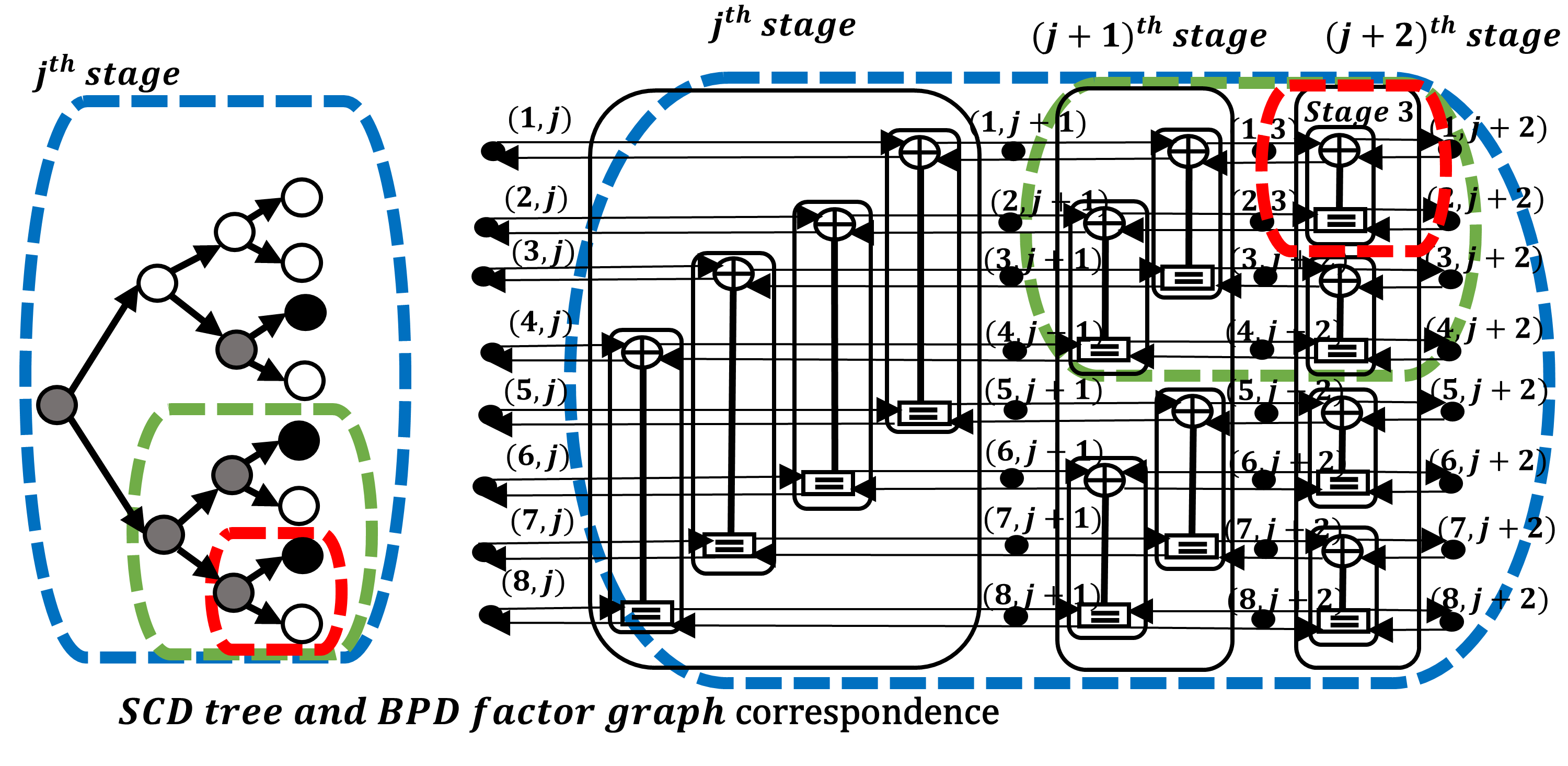}\protect\caption{SCD scheduling tree and factor graph of the (8,4) polar code}
\end{figure}

\begin{figure}[!th]
\includegraphics[width=0.9\columnwidth]{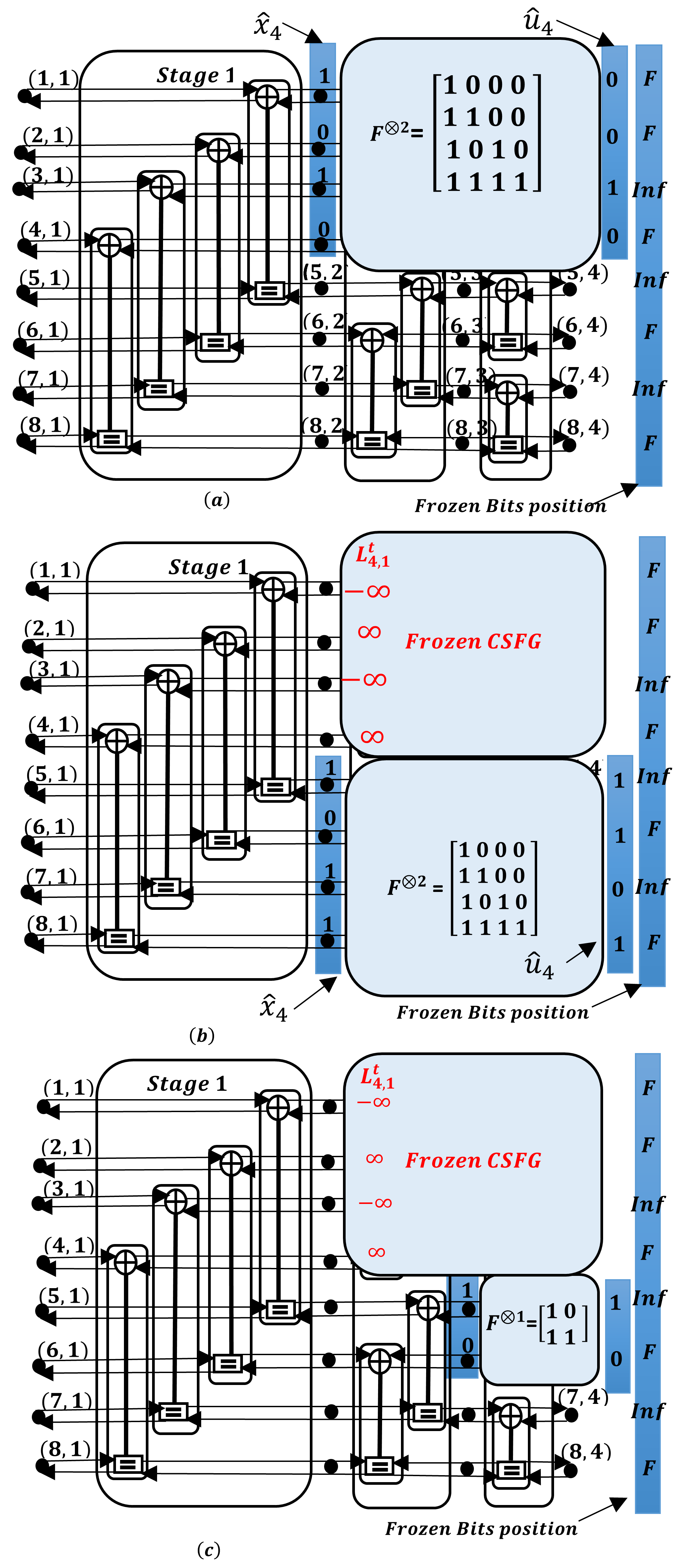}

\protect\caption{Example illustrating the Proposed Scheme (a) Checking first CSFG at
stage 1 (b) Checking second CSFG at stage 1 (c) Checking third CSFG
at stage 2}
\end{figure}

\section{THE PROPOSED SCHEME}

Fig. 3(a) shows the scheduling tree of the successive cancellation
decoding (SCD) of the (8,4) polar code \cite{jason1}, and Fig. 3(b)
depicts the equivalent BPD factor graph of the same (8,4) polar code.
At each stage the SCD scheduling tree is split into a number of sub-trees,
each of which is responsible for decoding a corresponding constituent
code. The size of the sub-tree varies at each level and is reduced
by half when moving from one stage to another stage. 

Before presenting the details of our proposed scheme, we first introduce
the notion of the \emph{connected sub-factor-graph}. A connected sub-factor
graph (CSFG) is defined as a sub-factor-graph which has the same number
of inputs and outputs and where the output nodes are at the stage
$m+1$ and each input is connected to each output through some PEs
in the sub-factor-graph. Fig. 3(b) shows two examples of CSFGs. It
can be seen that each CSFG has a corresponding sub-tree in the scheduling
tree of SCD. Fig. 3 (a) and (b) show examples of the corresponding
sub-trees and the connected sub-factor-graph of the (8,4) polar code.
The number of CSFGs at each stage is given by $2^{j}$, where $j$
is the stage number. For the (8,4) polar code, as shown in Fig. 3
(c) and (d), the numbers of CSFGs at stages 1 and 2 are 2 and 4, respectively. 

At each iteration $t$, the nodes at stage $j$ in the BPD factor
graph output left-to-right LLR-based propagating messages $R_{1:n,j+1}^{t}$,
and these are the inputs to the $2^{j}$ CSFGs at stage $j$. $R_{1:2^{m-j},j+1}^{t}$
are the inputs to the first CSFG, while $R_{((k-1)2^{m-j}+1):(k2^{m-j}),j+1}^{t}$are
those for the $k^{th}$ CSFG . Each CSFG is responsible for the decoding
of the corresponding constituent code from its respective input messages. 

The proposed scheme borrows the idea of successive cancellation decoding
(SCD), where the results of the previous-decoded bits are used for
the decoding of the current bit. Here we introduce a CSFG freezing
concept for a low complexity BPD. At a particular iteration $t$,
when the message passing reaches a certain stage $j$, if a CSFG at
that stage can correctly decode its corresponding constituent code
(i.e. the CSFG has reached convergence), it is frozen and no message
passing or updating within the CSFG will be needed in the subsequent
iterations. The details of how to check whether a CSFG can be frozen
will be presented later. 

One important thing is the checking order for the freezing of the
CSFG. A CSFG can only be frozen if all the previous CSFGs (in the
order of the decoding bits) at that stage have been frozen. If a CSFG
is not frozen, it means the message values inside it will still be
changed in the subsequent iterations. Similar to the SCD operation,
the message values of this CSFG will be used for the decoding of the
constituent codes of the subsequent CSFGs. Therefore the freezing
of the CSFGs at a stage has to follow an order based on the decoded
bit. When a CSFG at a certain stage is checked for freezing, if it
cannot correctly decode its constituent code, then it cannot be frozen
and the message passing and updating have to be executed for PEs at
that stage. After that, we move to the next stage and check the convergence
of the corresponding CSFGs. When we move to the next stage, the number
of CSFGs will be doubled. This freezing-checking procedure will continue
from stage to stage until the end of the BPD factor graph is reached. 

Next we will present how we can freeze a CSFG. As discussed above,
a CSFG corresponds to a sub-tree in the SCD scheduling tree, which
can also be viewed as a constituent code of the original polar code.
At the $t^{th}$ iteration and stage $j$, the left-to-right propagation
messages $R_{((k-1)2^{m-j}+1):(k2^{m-j}),j+1}^{t}$ connected to the
$k^{th}$ CSFG can be viewed as the LLR inputs to decode the corresponding
constituent code. We can apply Maximum-Likelihood Decoding (MLD) on
this constituent code with $R_{((k-1)2^{m-j}+1):(k2^{m-j}),j+1}^{t}$as
input to obtain a decoded output vector $(u_{((k-1)2^{m-j}+1):(k2^{m-j})})$,
which is a sub-vector of the source word ($\mathbf{u}_{n}$) of the
original polar code. As will be shown later, if the freezing of the
CSFGs follows the proposed order, the input messages of CSFG $R_{((k-1)2^{m-j}+1):(k2^{m-j}),j+1}^{t}$
are reliable enough and MLD $(u_{((k-1)2^{m-j}+1):(k2^{m-j})})$,
based on these input messages, can be taken as the decoded result
of the constituent code. The freezing order of the CSFG has to follow
the decoded bit order, and the top CSFGs at each stage will be frozen
first.

Fig. 4 shows the SCD scheduling tree and the factor graph of the (8,4)
polar code. We can see that the top CSFGs are actually corresponding
to the first few sub-trees that follow the depth-first traversal of
the SCD scheduling tree. At the first iteration, the input messages
to these CSFGs are the same as the input LLR messages of the corresponding
SCD sub-trees. Hence if we can decode the input messages of these
CSFGs using MLD, the decoding performance on the corresponding constituent
code will achieve or even exceed that of SCD. If the CSFGs cannot
be frozen at this iteration, and need further iteration to converge,
due to the nature of the iterative decoding, the reliability of the
input messages to these CSFGs will become better and hence the input
LLR messages of these CSFGs will be more reliable than the input messages
to the SCD sub-tree. As a result the MLD performance will not be worse
than that of SCD.

\begin{figure*}[tp]
\subfloat[\label{fig:ErrorPerformance}]{

\protect\includegraphics[width=1\columnwidth]{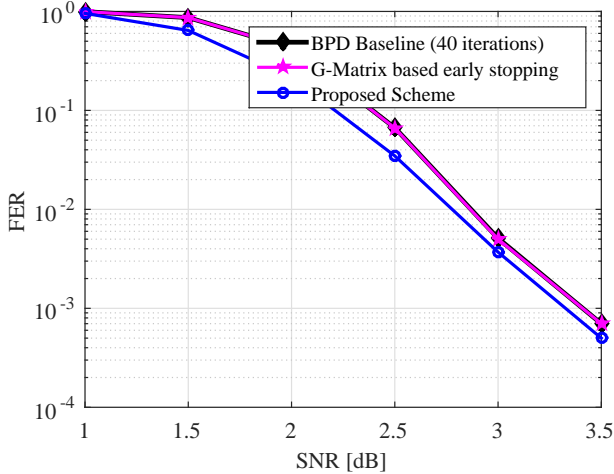}}\subfloat[\label{fig:iterations}]{

\protect\includegraphics[width=1\columnwidth]{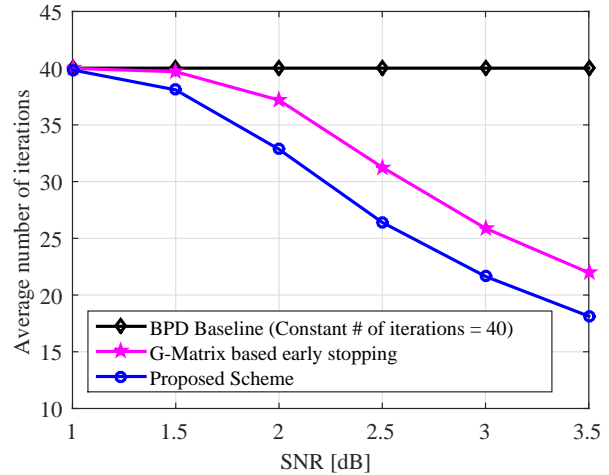}}

\subfloat[\label{fig:computations} ]{

\protect\includegraphics[width=1\columnwidth]{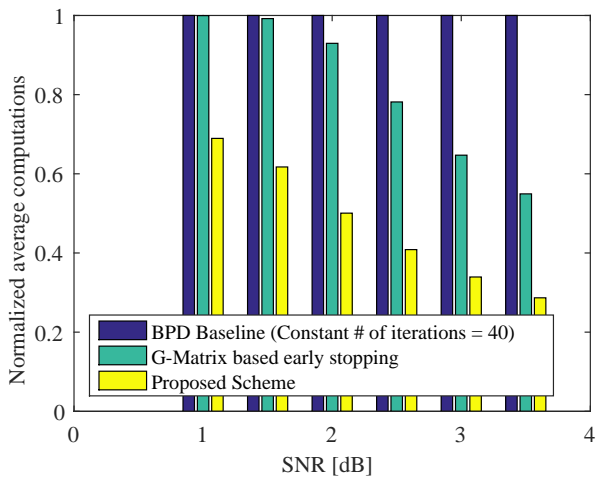}}\subfloat[\label{fig:savings} ]{

\protect\includegraphics[width=1\columnwidth]{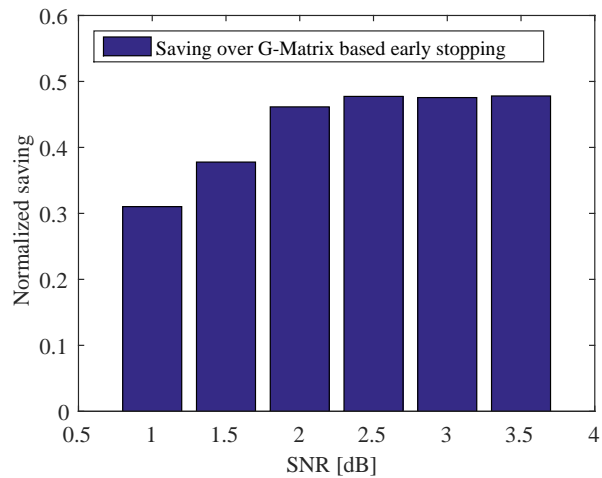}}

\protect\caption{\label{results}Comparison results for a (1024,512) polar code (a)
Error correction performance (b) Average number of iterations (c)
Average required computations (d) Computations savings over G-Matrix
based early stopping}
\end{figure*}

MLD is based on an exhaustive search and hence it has a huge complexity.
To reduce the complexity, novel checking criterion is suggested to
efficiently find the MLD result of the constituent code. Let $R_{1:2^{m-j},j+1}^{t}$
be the left-to-right propagation messages of a CSFG at stage $j$.
We obtain a hard decision vector $\hat{\mathbf{x}}{}_{2^{m-j}}=[\hat{x}{}_{1}\ldots\hat{x}{}_{2^{m-j}}]$
for these messages where

$\hat{\mathbf{x}}_{2^{m-j}}=\begin{cases}
\begin{array}{c}
0\\
1
\end{array} & \begin{array}{c}
\textrm{if}\left.\right.R_{1:2^{m-j},j+1}\geq0\\
\textrm{if}\left.\right.R_{1:2^{m-j},j+1}<0
\end{array}\end{cases}\left.\left.\left.\left.\left.\left.\left.\left.\left.\left.\right.\right.\right.\right.\right.\right.\right.\right.\right.\right.(2)$

Given $\hat{\mathbf{x}}_{2^{m-j}}$ as input to the CSFG, the decoded
bit vector at its output $\hat{\mathbf{u}}_{2^{m-j}}$, which is also
a sub-vector of the source word of the original polar code $\hat{\mathbf{u}}_{n}$,
is obtained by the inverse operation of polar code encoding that is
given as 

$\hat{\mathbf{u}}_{2^{m-j}}^{T}=\hat{\mathbf{x}}_{2^{m-j}}^{T}(F^{\otimes(m-j)})^{-1}=\hat{\mathbf{x}}_{2^{m-j}}^{T}(F^{\otimes(m-j)}),\left.\left.\left.\left.\left.\left.\left.\left.\left.\left.\right.\right.\right.\right.\right.\right.\right.\right.\right.\right.(3)$

$where\left.\right.(F^{\otimes(m-j)})^{-1}=(F^{\otimes(m-j)})$

Fig. 5(a) shows an example of hard decision decoding. The CSFG can
be frozen if the sub-source-word vector $\hat{\mathbf{u}}_{2^{m-j}}$
satisfies the following frozen set criteria: 

$u_{k}=0,\left.\right.for\left.\right.k\in\mathcal{A^{C}}\left.\left.\left.\left.\left.\left.\left.\left.\left.\left.\right.\right.\right.\right.\right.\right.\right.\right.\right.\right.(4)$

The following lemma shows that if the frozen set criteria (4) are
satisfied, the sub-source-word vector $\hat{\mathbf{u}}_{2^{m-j}}$
obtained by (3) is indeed the decoding results of the MLD on the corresponding
constituent code of the CSFG. 
\begin{lem}
Let $R_{1:2^{m-j},j+1}$and $\hat{\mathbf{x}}_{2^{m-j}}$ be the input
LLR messages and hard decision vector based on (2) for the corresponding
CSFG at the $j^{th}$ stage. If $\hat{\mathbf{u}}_{2^{m-j}}$ is obtained
from $\hat{\mathbf{x}}_{2^{m-j}}$ based on (3) and it satisfies the
frozen-set criteria of (4), then $\hat{\mathbf{u}}_{2^{m-j}}$ is
the maximum likelihood detection (MLD) result of the corresponding
constituent code with input messages $R_{1:2^{m-j},j+1}$.\end{lem}
\begin{IEEEproof}
The CSFG at the $j^{th}$ stage represents a short polar (constituent)
code of length $2^{m-j}$. Its input and output are related by $\mathbf{x}_{2^{m-j}}=\mathbf{u}_{2^{m-j}}F^{\otimes(m-j)}$.
From \cite{yazdi} and \cite{21}, given the input LLR $R_{1:2^{m-j},j+1}$,
the likelihood value of an arbitrary source word $u_{2^{m-j}}$ is
given by $\sum_{i=1}^{2^{m-j}}(1-2x_{2^{m-j}}[i])R_{1:2^{m-j},j+1}[i]$,
where $\mathbf{x}_{2^{m-j}}=\mathbf{u}_{2^{m-j}}F^{\otimes(m-j)}$. 

If no source word bit is a frozen bit, i.e., $u_{i}$ can assume both
0 and 1 for $1\leq i\leq2^{m-j}$, the source word $\hat{\mathbf{u}}_{2^{m-j}}$
obtained from $\hat{\mathbf{x}}_{2^{m-j}}$ has a maximum likelihood
value which is equal to $\sum_{i=1}^{2^{m-j}}|R_{1:2^{m-j},j+1}[i]|.$
If a certain source word bit is a frozen bit, the searching space
of the valid source word is smaller and $\sum_{i=1}^{2^{m-j}}|R_{1:2^{m-j},j+1}[i]|$
may not be achieved. However, if $\hat{\mathbf{u}}_{2^{m-j}}$ satisfies
(4), this likelihood value is achievable and the source word $\hat{\mathbf{u}}_{2^{m-j}}$
is a valid source word. Hence, $\hat{\mathbf{u}}_{2^{m-j}}$ is the
MLD result.
\end{IEEEproof}
When a CSFG at stage $j$ is frozen, the corresponding computations
and message updating are not needed for the rest of the iterations.
We can also fix its right-to-left feedback propagating messages $(L_{1:2^{m-j},j+1}^{t})$
for the rest of the iterations based on its $\hat{\mathbf{x}}_{2^{m-j}}$
since the output decoding decision for this CSFG has already been
made and we have

$L_{1:2^{m-j},j+1}^{t\in\{t,t+1,...\textrm{max\_iter}\}}=(-1)^{\mathbf{\hat{x}}_{2^{m-j}}}\infty$

In one iteration, propagating messages from left to right, for any
CSFG, if the frozen set criteria (4) is not satisfied then we cannot
freeze this CSFG. We then update the messages at this stage using
equation (1), move to the next stage and repeat the same procedure.
Fig. 5(b) shows an example. At the second iteration, we check the
bottom CSFG at stage 1. $\hat{\mathbf{u}}_{4}$ does not satisfy the
frozen bit criteria (4) and we cannot freeze this CSFG. So the messages
are updated at stage 1 and we move to the next stage (stage 2) to
check whether the first un-frozen CSFG can be frozen at stage 2, as
shown in Fig. 5(c). A CSFG can only be considered for freezing if
all the preceding CSFGs at the same stage have been frozen. This procedure
is repeated until all the CSFGs at a stage are frozen or we reach
the maximum number of iterations, which corresponds to the completion
of the decoding process. 

With the freezing of CSFGs, computations and message updating operations
do not need to be executed for rest of the iterations. Therefore the
overall computation complexity, and hence the energy consumption,
are reduced. Moreover the right-to-left feedback propagating messages
$(L_{1:2^{m-j},j+1}^{t})$ are fixed to either -\ensuremath{\infty}
or +\ensuremath{\infty} depending on the value of the hard-decision
bit when a CSFG is frozen. This boosts the reliability of the feedback
messages and will help the rest of the unfrozen CSFGs to converge
faster in the subsequent iterations, thus helping to reduce the overall
number of iterations for the decoding and hence the average latency.

\section{SIMULATION RESULTS}

To verify the error correcting performance and complexity saving for
the proposed frozen-CSFG-based BPD scheme, we carry out a simulation
on a polar code of length 1024 and rate \textonehalf{} and compare
the result with the original BPD scheme in \cite{14} (which we denote
as the baseline BPD) and the BPD using a G-matrix-based stopping criterion
in \cite{18}. Fig. 6 shows the simulation results over an AWGN channel
with BPSK modulation. For a fair comparison, we use the same set of
parameters as \cite{18}, where min-sum approximation with scaling
parameter $(\alpha=0.9375)$ and max\_iter = 40 were used. As seen
in Fig. 6(a), the proposed method has no performance degradation compared
with the other two existing BPD schemes. The average number of iterations
required for decoding a code word are compared in Fig. 6(b). It can
be seen that the proposed method requires the least number of iterations,
resulting in lower latency and higher throughput compared to \cite{18}.
At SNR = 3dB, the average number of iterations is reduced by 46\%
and 17\% when compared to the baseline BPD and the G-matrix-based
early stop method, respectively.

We also compare the overall computation complexity of the three BPD
schemes. For each PE in the factor graph, we count the number of iterations
until its operation is frozen in the proposed scheme. We then sum
the number of iterations, for which that PE is active, for all the
PEs. For the other two schemes, since every PE needs to be executed
in every iteration, the computation complexity just depends on the
average number of iterations.

Fig. 6(c) shows the normalized average number of computations required
for all three schemes. It can be observed that the proposed scheme
requires the least number of computations, which translates directly
to lower power consumption and latency for the overall decoding process.
It can be seen that at SNR = 3dB, the average computation complexity
is reduced by 65\% and 46\% when compared with the baseline scheme
and the early-stopping scheme, respectively. As state of the are BPD,
computaion savings for the proposed method are comaperd with G-matrix
based early stopping method in Fig. 6(d).

\section{CONCLUSION }

In this work we have presented a novel scheme to reduce the average
number of computations as well as average latency in belief propagation
decoding (BPD) for polar codes based on the concept of a frozen connected
sub-factor-graph. Simulation results show that there is no performance
degradation of the proposed scheme when compared with the original
belief propagation algorithm and the G-matrix-based early stopping
criterion, while the scheme enjoys a 46 \textasciitilde{} 65 \% reduction
in computation complexity, and 17 \textasciitilde{} 46\% reduction
in latency at SNR = 3dB. In future work, the VLSI architecture and
a hardware implementation will be developed.

\end{document}